\documentclass[12pt]{article} 
\usepackage[top=30mm, bottom=40mm, left=25mm, right=25mm]{geometry}
\usepackage[usenames,dvipsnames]{xcolor}
\usepackage{WDL3}

\usepackage{mathtools}

\newcommand{\ha}{{\hat{a}}}
\newcommand{\hb}{{\hat{b}}}
\newcommand{\hc}{{\hat{c}}}

\newcommand{\willsymbol}{\Pisces}
\newcommand{\stephensymbol}{\Leo}
\renewcommand{\epsilon}{\varepsilon}

\title{Superspace de Rham Complex and Relative Cohomology}
\author{William D. Linch \textsc{iii}$^\text{\willsymbol}$ and Stephen Randall$^\text{\stephensymbol}$}
\date{}

\begin{document}
\maketitle

\begin{center}
{\em
Center for String and Particle Theory\\
Department of Physics,
University of Maryland at College Park,\\
College Park, MD 20742-4111 USA.
}
\end{center} 

\vspace{10pt}

\begin{abstract}
We investigate the super-de Rham complex of five-dimensional superforms with $N=1$ supersymmetry. By introducing a free supercommutative algebra of auxiliary variables, we show that this complex is equivalent to the Chevalley-Eilenberg complex of the translation supergroup with values in superfields. Each cocycle of this complex is defined by a Lorentz- and iso-spin-irreducible superfield subject to a set of constraints.
Restricting to constant coefficients results in a subcomplex in which components of the cocycles are coboundaries while the constraints on the defining superfields span the cohomology. This reduces the computation of all of the superspace Bianchi identities to a single linear algebra problem the solution of which implies new features not present in the standard four-dimensional, $N=1$ complex. These include splitting/joining in the complex and the existence of cocycles that do not correspond to irreducible supermultiplets of closed differential forms.
Interpreting the five-dimensional de Rham complex as arising from dimensional reduction from the six-dimensional complex, we find a second five-dimensional complex associated to the relative de Rham complex of the embedding of the latter in the former. This gives rise to a second source of closed differential forms previously attributed to the phenomenon called ``Weyl triviality''. 
\end{abstract}

\vspace*{.5cm}
\begin{flushleft}
~\\
{${}^{\small \mbox\willsymbol}$ \href{mailto:wdlinch3@gmail.com}{{wdlinch3@gmail.com}}}\\
{$^\text{\stephensymbol}$ \href{mailto:stephenlrandall@gmail.com}{{stephenlrandall@gmail.com}}}
\end{flushleft}

\setcounter{page}0
\thispagestyle{empty}

\newpage

\tableofcontents

\section{Introduction}

The systematic study of closed differential forms in superspace began with the work of \cite{Gates:1980ay} on four-dimensional $p$-forms with four supercharges. Since then much effort has gone into the construction of such forms in various dimensions with various amounts of supersymmetry. In four dimensions with $N=1$ supersymmetry, this has been a textbook subject for some time now \cite{Gates:1983nr}. The analogous study of closed forms in four dimensions with $N=2$ supersymmetry was performed in harmonic superspace in \cite{Biswas:2001wu}. 
The super-de Rham complex in five dimensions with $N=1$ supersymmetry is presented in \cite{Gates:2014cqa} 
while that in six-dimensional curved, $N=(1,0)$ superspace was constructed in \cite{Arias:2014ona}. 
In addition to these attempts at systematic studies, results on specific such forms in superspace with and without (gauged) central charges ({\it e.g.} \cite{Hindawi:1996gd, Buchbinder:1997pe,Butter:2012ze}) and their application to supersymmetric field theory ({\it e.g.} \cite{Dragon:1997za,Gates:1980az,Bergshoeff:1996qm,Bandos:2013sia, Kuzenko:2013rna,Buchbinder:2014sna}), and gravity ({\it e.g.} \cite{Gates:1981qq,Gates:2009xt,Novak:2012xz})
are scattered throughout the literature. 

In many of these studies, one is struck by the effort required to obtain the superfield description of the $p$-form and the complexity of the structure of its components, even in flat superspace. After all, the analogous problem in the theory of smooth manifolds is solved universally by the Poincar\'e Lemma: Any closed $p$-form $\omega$ on a contractible space is the exterior derivative of a $(p-1)$-form $\eta$, or $d\omega = 0$ $\Rightarrow$ $\omega = d\eta$. In the case of superforms, this solution is unacceptable when given in local supercoordinates $\{z^M\}$ as $\partial_{[M_1}\eta_{M_2\dots M_p]}(z)$ because these components are not superfields. As we review below, one remedies this by passing to a description in terms of frames and superspace covariant derivatives. This complicates the problem because the former carry torsion and the latter are not differentials. 
Nevertheless, the solution in superspace should not be as complicated as suggested by perusal of the literature on the subject given that, in the end, it is just $d\eta$.

In this paper we attempt to show that the complexity of the standard calculational method is due primarily to a redundancy in the analysis that can be avoided by carefully separating the constraints on the superfields defining the components of the form. 
The problem of separation is solved automatically
when the components of the form are interpreted as cocycles in an algebraic differential complex (closely associated to the de Rham complex) that can be thought of as encoding certain Fierz identities. 
Even the calculation of the Fierz identities can be avoided almost completely because the only relevant ones follow immediately from a famous $\gamma$-matrix identity valid in dimensions $D=2^k+2$ (for appropriate spinor representations).
Taken together, the computation of the components of the form and the constraints on its defining superfield is reduced dramatically.\footnote{In this work, we will not attempt to solve (in terms of prepotentials) the constraints arising on the superfields that define the forms as such solutions are well-known in these cases. With this understood, by ``solving'' the Bianchi identities for a form $\omega$, we will mean only that we have found the components of $\omega$ in terms of a specific field strength superfield $\phi$ and that we have found all of the constraints to which $\phi$ is subject.} 

Besides the practical aspect of reducing the work required to find the explicit structure of closed differential forms in superspace, this interpretation of the problem elucidates certain generic properties of the complices of super-cocycles in superspace. For example, we will show that generally the complex 
will have loops (branching and fusion) and that some of its $p$-cocycles are not the supersymmetric generalization of closed differential forms.

The interpretation we advocate in this paper applies to all superspaces, provided the appropriate modifications are made. In order to avoid an overly-formal analysis, however, we have opted to present the construction by focusing in detail on the case of flat five-dimensional superspace.\footnote{A complete analysis of this complex is given presented in reference \cite{Gates:2014cqa} without the use of the machinery introduced here.} In doing so, however, we have used only those techniques that apply to flat superspaces (without central charge) in any dimension. With this approach, we hope to have succeeded in striking a balance between application and theory by explicitly demonstrating the implementation of the method 
on a few examples while abstaining, where possible, from the use of case-specific methods.

\paragraph{Outline} We have structured the presentation as follows. In section \ref{S:General} we begin with the textbook definition of differential superforms. Following reference \cite{Arias:2014ona}, we then introduce a supercommutative algebra of auxiliary variables that allow us to recast the super-de Rham complex into a calculationally more useful form. The resulting complex turns out to be the Chevalley-Eilenberg complex of the supersymmetry algebra with superfield coefficients. This complex admits the action of a second differential (not commuting with the Chevalley-Eilenberg differential) with respect to which the coefficients of the cocycles are coboundaries and the constraints on the superfields that define it are in cohomology. 

We then specialize to five-dimensional, $N=1$ superspace in section \ref{S:5DN1deRham} and explicitly show how this reinterpretation of the de Rham complex is used to determine the components of the cocycles and the constraints on their defining superfields for the cases $p=1$, $2$, and $3$. In the process, we discover that the complex splits and rejoins in the transition $1\to 2\to 3$ thereby creating a loop (cf. fig. \ref{F:Branching}). We also find that certain $p$-cocycles are not supersymmetric versions of $p$-forms with, for example, the $3$-cocycle having the interpretation of a multiplet of superconformal gauge transformation parameters, instead. 

We then switch gears in section \ref{S:RelativeCohomology} and discuss the embedding of the supersymmetric de Rham complex in six-dimensional, $N=(1,0)$ superspace. Reducing back down to five dimensions, we find a second complex related to the supersymmetric version of the relative de Rham complex \cite{bott1995differential} of the embedding of the five-dimensional space into the six-dimensional one. Here we find the missing closed $3$-form and comment on its relation to ``ectoplasm with an edge'' \cite{Howe:2011tm} and ``Weyl triviality'' \cite{Bonora:1986xd}.
 
We conclude in section \ref{S:Outlook} with a few comments regarding the interpretation and generalization of our results and their applications to open problems in superspace. In appendix \ref{S:Superspace} we work out the two main formul\ae{} generating the cohomology of the five- and six-dimensional complices. We do this in a way that generalizes to any superspace that can be embedded as a hypersurface in a ``principal'' superspace in which the pairing (spinor)$\otimes$(spinor)$\to$(vector) of commuting spinors to make a vector is null. Presented in this way, the analysis may be carried over to superspaces of other dimensions.


\section{General Setup}
\label{S:General}
A super-$p$-form $\omega$ is given in local coordinates by the formula \cite{Gates:1983nr,Buchbinder:1998qv,Wess:1992cp}
\begin{align}
\omega = \frac1{p!} dz^{M_1} {\wedge} \dots  {\wedge} dz^{M_p} \omega_{M_p \dots M_1}(z)
\end{align}
with the collection of forms acted on by the supersymmetric analogue of the de Rham differential $d=dz^M\partial/\partial z^M$. Such a form is ``closed'' when $d\omega = 0$ and ``exact'' when $\omega = d\eta$ for some form $\eta$ of degree 1 lower. The super-analogue of the Poincar\'e lemma states that any closed form on a superspace with contractible body is exact so the solution to the condition $d\omega =0$ 
is $\omega = d\eta$ where $\eta$ is only defined up to a redefinition by a closed term. 

This solution, however, is not a superfield representation of supersymmetry because the odd part of the super-de Rham differential does not commute with the supersymmetry generators, even in flat space. To remedy this, one passes to an invariant framing
\begin{align}
d = dz^M \partial_M = e^A D_A,
\end{align}
where $\{e^A\}$ denotes a basis of the left-invariant 1-forms of the super-translation group and $D_A$ are the supercovariant derivatives that commute in the graded sense with translations. In this description, the form is re-expressed as 
\begin{align}
\omega = \frac1{p!} e^{A_1} {\wedge} \dots {\wedge} e^{A_p} \omega_{A_p \dots A_1}(z).
\end{align} 
and its components $\{\omega_{A_p\dots A_1}(z)\}$ are all superfield representations of the translation superalgebra (supersymmetry algebra). The price to pay for this is that the frames carry non-zero torsion 
\begin{align}
de^A = T^A 
\end{align}
even in flat superspace, and the equations for the components 
of a closed form become the ``Bianchi identities'' 
\begin{align}
\label{E:BianchiID}
{1\over (p+1)!}(d\omega)_{A_1 \dots A_{p+1}} =
	{1\over p!} D_{[A_1} \omega_{A_2 \dots A_{p+1}]} 
	+ {1\over 2! (p-1)!}  T_{[A_1 A_2|}{}^C \omega_{C|A_3 \dots A_{p+1}]} = 0.
\end{align}

\subsection{The Chevalley-Eilenberg Complex} 
\label{S:CE}
In reference \cite{Arias:2014ona} it was advocated that the presentation and analysis of the super-de Rham complex in six-dimensional, $N=(1,0)$ superspace is simplified by the introduction of a supercommutative set of variables $s^A$ to replace the frames $e^A$.\footnote{The utility of such variables goes far beyond this by aiding in the identification of certain integrable distributions that, in turn,
simplify the classification of superconformal field representations and assist in the construction of supersymmetric integration measures. We do not address this further in this work (but see {\it e.g.} ref. \cite{Arias:2014ona} for the relation to six-dimensional curved projective superspace \cite{Linch:2012zh}).
} 
In contrast to the frames, the new variables are constants $ds^A = 0$ so that by replacing $e\to s$ everywhere, we are constructing a new complex in which the differential is no longer the original de Rham differential and the torsion must be treated separately. Splitting the $s^A$ variables into a spinor part, denoted by $s$, and a vector part, denoted by $\psi$, differential forms fan out into a collection of objects 
\begin{align}
\omega_{\unbr{s\dots s}s \unbr{\psi\dots \psi}{p-s}}
	= s^{\alpha_1} \dots s^{\alpha_s} \psi^{a_1}\dots \psi^{a_{p-s}} 
		\omega_{\alpha_1 \dots \alpha_s a_1\dots a_{p-s}}
\end{align}
graded by number of $s$s (and total degree $p$).

For the sake of clarity of exposition we now specialize to flat space.\footnote{We comment on the generalization to curved superspace in section \ref{S:Outlook}.}
Then the collection of components is acted on by the graded derivations $D_s = s^\alpha D_\alpha$ and $\partial_\psi = \psi^a \partial_a$ which satisfy the flat-space covariant derivative algebra rules expressed succinctly by the single non-trivial relation
\begin{align}
D_s^2 = i \partial_{\gamma(s,s)} .
\end{align}
Here and throughout, we employ a compact notation in which indices contracted with an object are labelled by that object, and $\gamma(s,s)$ stands for the vector $s^\alpha s^\beta (\gamma^a)_{\alpha \beta}$ so that, for example, $\partial_{\gamma(s,s)}$ is the combination $s^\alpha s^\beta (\gamma^a)_{\alpha \beta} \partial/\partial x^a$.\footnote{In five and six dimensions, the spinor representation used is pseudo-real and the Pauli matrices are anti-symmetric so that $\partial_{\gamma(s,s)}$ really stands for $s^{\alpha i} s^{\beta j} \varepsilon_{ij} (\gamma^a)_{\alpha \beta}$. Such nuances are not important for our exposition so we will suppress them throughout this section (but see appendix \ref{S:Superspace}).
} 

In the new complex, the differential of a form $\{\omega_{s\dots s\psi \dots \psi}\}$ is defined by the collection of expressions
\begin{align}
\label{E:BianchiForm}
B(\omega)_{\unbr{s\dots s}{s+1} \unbr{\psi \dots\psi}{p-s} } := 
	(s+1)D_s \omega_{s \dots s \psi \dots \psi} 
	-(-1)^s (p-s) \partial_\psi \omega_{s \dots s \psi \dots \psi} 
	+ i (-1)^s s(s+1) \omega_{s \dots s \gamma(s,s)\psi \dots \psi} ,
\end{align}
where $s$, when used as a coefficient, stands for the number of spinor variables $s^\alpha$ in the formula save one.
Note that these are proportional to the components of the Bianchi identities (\ref{E:BianchiID}) with $s$ and $\psi$ variables contracted. That is, $B(\omega)_{s\dots s\psi\dots \psi} \propto (d\omega)_{s\dots s\psi \dots \psi}$. That the map $\omega \mapsto B(\omega)$ is a differential follows from the ``Bianchi identity for Bianchi identities'' 
\begin{align}
\label{Bianchi4Bianchi}
(s+1)D_s B_{s \dots s \psi \dots \psi} 
	-(-1)^s (p-s) \partial_\psi B_{s \dots s \psi \dots \psi} 
	+ i (-1)^s s(s+1) B_{s \dots s \gamma(s,s)\psi \dots \psi} =0 ,
\end{align}
which follow from $B(B(\omega))_{s\dots s\psi\dots \psi} \propto B(d\omega)_{s\dots s\psi \dots \psi} \propto (dd\omega)_{s\dots s\psi \dots \psi} \equiv 0$.

We claim that $B$ (considered as a map $\omega \mapsto B(\omega)$) is equivalent to the Chevalley-Eilenberg differential $d_{CE}$ \cite{Chevalley:1948zz} for the superalgebra of odd and even translations generated by $Q$ and $P$, respectively.\footnote{This observation is due to Paul Green.} 
The latter is defined on a complex with a basis freely generated by the $s$ and $\psi$ variables. Then $d_{CE} = P_\psi + Q_s +\dots$ where the corrections are terms proportional to the structure constants of the Lie superalgebra that ensure that $d_{CE}^2 =0$ on Lie algebra {co}cycles. This uniquely determines $d_{CE} = Q_s + P_\psi +  \iota_{\gamma(s,s)}$ where $\iota_v \omega_{s\dots s \psi\psi\dots \psi} = \omega_{s\dots s v\psi\dots \psi}$ denotes contraction by the vector $v$. In particular, by the supersymmetry algebra,
\begin{align}
\{ Q_s, Q_s \} = - 2P_{\gamma(s,s)} 
~~~\Rightarrow~~~
d_{CE}^2 = Q_sQ_s + P_{\gamma(s,s)} = 0.
\end{align}

The action of the Lie superalgebra embeds in the super-vector fields on the supermanifold on which the superfields are defined and, thus, acts on the superfields as graded derivations. 
Although it is conventional to define the Chevalley-Eilenberg differential by the action of the generators of the Lie algebra, in our case it is more convenient (and equivalent) to define the action on the module of superfields by the covariant derivatives instead. Thus, we conclude that the super-de Rham complex is equivalent
to the Chevalley-Eilenberg complex for the supersymmetry algebra with values in the module of superfields.\footnote{Taken together with the conclusions reached in reference \cite{Arias:2014ona}, a version of this statement is expected to hold also for the Chevalley-Eilenberg complex in curved homogeneous superspaces. We defer discussion of this possibility to section \ref{S:Outlook}.
} 


The Chevalley-Eilenberg complex for the supersymmetry algebra has been investigated extensively by Brandt \cite{Brandt:2009xv,Brandt:2010fa,Brandt:2010tz,Brandt:2013xpa} who relates this cohomology to a reduced cohomology, as we do in the next section. In this approach, an obstruction theory is developed to check when a solution to the cohomology of the reduced complex lifts to a solution of the full complex. 
This analysis was extended by Movshev, Schwarz, and Xu to the super-Poincar\'e algebra in \cite{Movshev:2010mf, Movshev:2011pr}. 
In the next section we take a different approach that exploits the behavior of the Bianchi-for-Bianchi identities (\ref{Bianchi4Bianchi}) under contraction by the reduced differential.

\subsection{Reduction of Coefficients}
\label{S:AlgebraicComplex}

The conclusion reached in the previous section, while useful for theoretical purposes, does not, in itself, help us to {solve} the superspace Bianchi identities. For this, we introduce another complex. Rather, we recognize that (our version of) the Chevalley-Eilenberg complex already admits the action of a differential $\delta := \iota_{\gamma(s,s)}$ taking the $s|p$ component of a cocycle to the $(s+2)|(p-1)$ component of another cocycle.

Suppose we have a $p$-cocycle $\omega$ satisfying the condition that, for some $\ell$, the Bianchi identities
\begin{align}
B(\omega)_{\unbr{s\dots s}{p+1-q} \unbr{\psi \dots\psi}{q} } = 0 
\end{align}
hold for all $q\leq \ell$.
We will say that $\omega$ solves its Bianchi identities (or is closed) up to level $\ell$. (In particular, $\omega$ is closed iff it is solved up to level $\ell = p+1$.) Next, we observe that in the Bianchi for Bianchi identity (\ref{Bianchi4Bianchi}) the component with the highest number of bosonic indices is the last one and that the others have 1 or 2 fewer such indices (and, correspondingly, that many more spinor indices). Suppose then, that the $p$-cocycle $\omega$ is solved up to level $\ell$. Then equation (\ref{Bianchi4Bianchi}) implies that
\begin{align}
B(\omega)_{\unbr{s\dots s}{p-\ell}\gamma(s,s) \unbr{\psi \dots\psi}{\ell} } = 0 
~~~\Rightarrow~~~
B(\omega)_{\unbr{s\dots s}{p-\ell}\unbr{\psi \dots\psi}{\ell+1}} 
\in \mathrm{ker} \, \delta.
\end{align}
In other words, the ``next'' component of $B(\omega)$ is a {co}cycle of the new differential. 

From the algebraic standpoint, the space of components of a $p$-cocycle is an ordinary real vector space so that the space of all such cocycles splits into those that are annihilated by $\delta$ and those that are not. Let us denote by $Z$ the subspace of ones that are (the $\delta$-cocycles). In this language, we have just found that the ``level-$(\ell+1)$ component'' of the Bianchi form sits in $Z$. Since $\delta$ is a linear map, $Z$ itself splits into $Z=B\oplus H$ where $B:=\mathrm{im} \,\delta$ consists of coboundaries, and the cohomology $H= Z/B$ is its complement. 

Now consider the level-$(\ell +1)$ Bianchi components of a $p$-cocycle that has been solved up to level $\ell$:
\begin{align}
B(\omega)_{\unbr{s\dots s}{p-\ell} \unbr{\psi \dots\psi}{\ell+1} } = 
	(s+1)D_s \omega_{s \dots s \psi \dots \psi} 
	-(-1)^s (p-s) \partial_\psi \omega_{s \dots s \psi \dots \psi} 
	+ i (-1)^s s(s+1) \omega_{s \dots s \gamma(s,s)\psi \dots \psi} .
\end{align}
This expression, again, splits into $B\oplus H$, and it is clear that the last term is entirely in $B$. Splitting this equation thus, there is a part of the first two terms that sits in $B$ while the rest sits in $H$. When we solve this Bianchi identity ({\it i.e.} set this component of $B(\omega)\to 0$), the terms in $B$ and $H$ must cancel separately ($B\cap H=\{0\}$) and we find that the next-level component (corresponding to the last term) is the part of the first two that sits in $B$. Furthermore, the remaining part, which sits in $H$ and must vanish separately, represents a condition on the lower components of $\omega$. That is, it is a {\em constraint} on these components.

Proceeding by induction on
the level $\ell = 0,\dots , p+1$, we see that the constraints on the components of the cocycles are determined at each level by the algebraic structure of $H$ while the definition of the components themselves are determined by that of $B$:
\begin{align}
\begin{array}{ccl}
\textrm{components of the $p$-cocycle} ~~~&\longleftrightarrow&~~~ B=\textrm{{co}boundaries of $\delta$} \cr
\textrm{constraints on ``solution''} ~~~&\longleftrightarrow&~~~ H=\textrm{cohomology of $\delta$} 
\end{array}.
\end{align}
With this, we have translated the superspace differential geometry problem of solving the Bianchi identities for a closed superform into an {\em algebraic} cohomology problem. 

\subsubsection{An Aside on the Mathematical Interpretation}

The construction in this section has a simple mathematical interpretation, which we describe briefly (and without pretense of mathematical rigor). 
First, we found in section \ref{S:CE} that the original super-de Rham complex for the superspace $M$ is equivalent to the Chevalley-Eilenberg complex $\Omega_{CE}^\bullet(\mathfrak g,\mathscr O_M)$ of the supersymmetry algebra $\mathfrak g$ of supertranslations of on $M$ with values in the $\mathfrak g$-module $\mathscr O_M$ of superfields on $M$. The differential $d_{CE}$ of this complex contains the differential $\delta$ introduced in this section and, in fact, reduces to it when the superfields on which it acts are constant. The conclusions reached in section \ref{S:AlgebraicComplex}, therefore, can be interpreted to mean that the problem of finding the closed super-de Rham forms and their constraints
reduces to the cohomology of the Chevalley-Eilenberg complex with constant coefficients tensored by the module of superfields on $M$. That is, very roughly speaking, 
$\Omega_{dR}^\bullet(M, \mathscr O_M) \sim 
\Omega_{CE}^\bullet(\mathfrak g,\mathscr O_M) 
\sim \mathscr O_M \otimes \Omega_{CE}^\bullet(\mathfrak g,\mathbb R)$. 
In the terminology of references \cite{Gates:1997kr,Gates:1998hy}, one could say that ``ectoplasm has no topology''. 
Note however, that in a more precise version of this formula, there will be a sum over non-trivial Lorentz and iso-spin representations 
in which the Chevalley-Eilenberg groups will take values \cite{Chevalley:1948zz}.

\section{Five-dimensional, $N=1$ super-de Rham Complex}
\label{S:5DN1deRham}
Our goal in this section is to apply the machinery we have proposed in section \ref{S:General} to derive the complex of closed super-de Rham forms \cite{Gates:1980ay,Gates:1983nr} to the case of flat five-dimensional, $N=1$ superspace \cite{Kuzenko:2005sz}. To do this, we need only the completeness relation\footnote{As explained in appendix \ref{S:Superspace}, the coefficients in the completeness relation (\ref{E:5DCompleteness}) are fixed by matching to the conventions established in reference \cite{Kuzenko:2005sz} and contain no important information. Nevertheless, these coefficients enter into the definition of the components and the explicit form of the constraints so they are needed for comparison to existing results.} 
\begin{align}
\label{E:bispinor}
X_s Y_s = 
	\tfrac{1}{8} s^2 X^iY_i
	- \tfrac{1}{8} (X^i \Gamma_{\Gamma(s,s)}Y_i) 
	+ \tfrac{1}{2} \Sigma^{\hat a \hat b}(s^i,s^j) (X_i \Sigma_{\hat a\hat b} Y_j),
\end{align}
for any two co-spinors $X$ and $Y$ and the cohomology of the complex with constant coefficients of section \ref{S:AlgebraicComplex}. In this superspace, this is generated entirely by the single non-trivial relation (cf. eq. \ref{E:5DCohomologyRelations})
\begin{align}
\label{E:CohomologyRelation}
\Gamma^\ha(s, s) \Sigma_{\ha \hb}(s^i, s^j) = 0 .
\end{align}
These two algebraic relations suffice 
to find the components and constraints of the closed super-de Rham forms in five-dimensional, $N=1$ superspace, as we now show.

\subsection{Closed Super-$1$-form}
\label{S:1cocycle}
A super-$1$-form $A$ is given in our complex as a pair $(A_s, A_\psi)$ subject to the vanishing of the Bianchi components (\ref{E:BianchiForm})\footnote{Note that in five dimensions the natural index contraction is $A^{\hat{\alpha}} B_{\hat{\alpha}}$ and so here we are defining $\Gamma^{\hat a}(s,\, s) := s^{\hat{\alpha} i} (\Gamma^{\hat a})_{\hat{\alpha}}{}^{\hat{\beta}} s_{\hat{\beta} i}$ which is off by a sign from the six-dimensional contraction $\gamma^a(s,\, s) = s^{\alpha} (\gamma^a)_{\alpha \beta} s^{\beta}$.}
\begin{align}
B_{ss} &= 2D_s A_s + 2i A_{\Gamma(s,s)} \cr
B_{s\psi} &= D_s A_\psi -\partial_\psi A_s\cr
B_{\psi\psi} &= \partial_\psi A_\psi.
\end{align}
Noting that $\delta B_{ss}$ is trivially 0 (so $B_{ss}\in Z$ is a {co}cycle), we now solve $B_{ss}=0$ by expanding out the first term using (\ref{E:bispinor}). The {co}boundary part (the part in $B=\mathrm{im}\delta$) is 
\begin{align}
- \tfrac{1}{4} D^i\Gamma_{\Gamma(s,s)} A_i + 2i A_{\Gamma(s,s)} = 0
~~~\Rightarrow~~~ A_\psi = -\tfrac{i}{8} D^i \Gamma_\psi A_i,
\end{align}
thereby defining the $A_\psi$ component in terms of $A_s$. 
Here we are using the facts that (i) we are allowed to solve $\delta(\dots)=0$ in $B$ up to a $\delta$-closed term $z_\psi$, and (ii) since there is no {co}cycle at this level ($\delta z\neq 0$ unless $z=0$), there is no such ``algebraic integration constant''. 

Now that we have removed the {co}boundary part $B$ from $Z=B\oplus H$, there is only the cohomology $H$ left. This part is given by the remaining terms
\begin{align}
s^2 D^iA_i + 4 \Sigma^{\hat a \hat b}(s^i,s^j) (D_i \Sigma_{\hat a\hat b} A_j) = 0 , 
\end{align}
which are linearly independent as bilinears in $s$. As such, we see that, at this level, $H=:H_{(1)}\oplus H_{(2)}$ has two parts with each giving an independent constraint
\hypertarget{C}{}
\begin{align}
\label{E:1formConstraints}
C:= D^i A_i = 0
~~~\mathrm{and}~~~
C_{\hat a \hat b ij} := D_{(i} \Sigma_{\hat a\hat b} A_{j)} = 0.
\end{align}
These equations can be solved in terms of some prepotential, 
but we will not need this solution explicitly here.\footnote{We are not claiming that the homological algebra procedure of section \ref{S:General} solves superspace constraints in terms of prepotentials. Rather, it merely {\em finds} these constraints for us. On the other hand, these superspace constraints are typically solved on some integrable subspace of the superspace augmented by the $s$ and $\psi$ variables ({\it i.e.} the superfield module in which the Chevalley-Eilenberg complex takes values). Famous examples 
include the chiral subspaces of ordinary superspaces \cite{Gates:1983nr,Buchbinder:1998qv}, the analytic subspaces of harmonic superspaces \cite{Galperin:1984av, Galperin:2001uw}, and the projective \cite{Karlhede:1984vr,Lindstrom:1987ks,Lindstrom:1989ne} and pure spinor superspaces \cite{Berkovits:2000fe} themselves. 
} 

We have ``solved'' the level-0 Bianchi identity in the sense that we have identified the superfield-strength ($A_s$) and the constraints that the level-0 Bianchi identity imposes on it (\ref{E:1formConstraints}). The level-1 identity has no {co}boundary part, consistent with the fact that there are no more components of $A$ left to determine. The cohomology at this level consists of all superfields of the form $F_{s\psi}$ that are annihilated by $\delta$ ({\it i.e.} a cocycle)
\begin{align}
F_{s\Gamma(s,s)} = 0
\end{align}
but are not {co}boundaries. Since {co}boundaries must have at least two $s$s, 
anything satisfying this equation is automatically in cohomology and, therefore, a constraint. However, the only identity that generates cohomology has 4$s$s (cf. eq. \ref{E:CohomologyRelation}) so that at this level, $H=\{0\}$, indicating that there are no more relations. This same reasoning applies {\it mutatis mutandis} to the level-2 Bianchi component. Therefore, our analysis of the closed super-1-form is complete.

\subsection{Closed Super-$2$-form}
\label{S:2cocycle}
We now repeat the analysis for the case of the closed super-$2$-form. The Bianchi components (\ref{E:BianchiForm}) are 
\begin{align}
B_{sss} &= 3 D_s F_{ss} - 6i F_{s\Gamma(s,s)} \cr
B_{ss\psi} &= 2 D_s F_{s\psi} + \partial_\psi F_{ss} + 2i F_{\Gamma(s,s)\psi} \cr
B_{s\psi\psi} &=  D_s F_{\psi\psi} - 2\partial_\psi F_{s\psi}  \cr
B_{\psi\psi\psi} &= 3\partial_\psi F_{\psi\psi} .
\end{align}
When we solve the level-0 identity, we encounter a new phenomenon: non-uniqueness of the solution. 
The forms in the de Rham complex have the property that they obstruct the closure of the previous form in the sequence. In this particular case, there are two choices corresponding to the obstruction of either of the two constraints in (\ref{E:1formConstraints}) and to proceed, we must select one of these two branches. 
We will revisit the non-uniqueness implied by the level-0 Bianchi identity in section \ref{S:Orogeny} but here we choose to present the analysis for the familiar case corresponding to the Lorentz and iso-spin singlet combination. That is, we take
\begin{align}
F_{ss} = 2is^2 W
~~~\mathrm{and}~~~
F_{s\psi} = -s^i\Gamma_\psi D_i W ,
\end{align}
for some scalar superfield $W$ with the normalization chosen to agree with \cite{Kuzenko:2005sz,Gates:2014cqa}. (Note that in five dimensions we must have $F_{ss} \neq 0$ due to the first equation in \ref{E:5DCohomologyRelations}.)

At level 1, the {co}boundary part defines the $F_{\psi\psi}$ component:
\begin{align}
- \tfrac{1}{2} D^i\Sigma_{\Gamma(s,s)\psi} D_iW + 2i F_{\Gamma(s,s)\psi}= 0
~~~\Rightarrow~~~
F_{\psi\psi} = -\tfrac{i}{4}D^i\Sigma_{\psi\psi} D_iW ,
\end{align}
where we have, again, used the fact that $\delta$ is uniquely invertible in $B$. With this component fixed, the remaining part(s) of the identity are in cohomology $H$ and define constraints. Having used up the $D^i\Sigma_{\Gamma(s,s)\psi} D_iW$ part in $DDW$, the remainder of the first term is a $\partial W$ term and a $(s\Sigma s) \cdot D\Gamma D W$ part. These are, again, in different irreducible representations of Lorentz and isospin symmetry and, since the $\partial W$ terms cannot form a {co}cycle by themselves (unless $W$ is constant), the partial terms must cancel. This leaves the term
\hypertarget{Caij}{}
\begin{align}
\label{E:Wconstraint}
s^i \Sigma_{\psi \ha} s^j D_{(i} \Gamma^\ha D_{j)} W = 0
~~~\Rightarrow~~~ 
C_{\hat a ij} := D^2_{\hat a ij} W = 0 ,
\end{align}
which is, indeed, the only cohomology (cf. eq. \ref{E:CohomologyRelation}) at this level. Here we have defined the shorthand $D^2_{\hat a ij}:=\frac12 D_{(i} \Gamma_\ha D_{j)}$ as this operator appears repeatedly in constraints and the definition of components.
The condition $D^2_{\hat a ij} W = 0$ is the defining constraint on the field strength superfield of the five-dimensional, $N=1$ vector multiplet \cite{Kuzenko:2005sz}. 

This completes the analysis of the level-2 Bianchi identity. In the previous subsection, we saw that there were no further conditions imposed at the next two levels (no cohomology there). Closer inspection of that argument implies that the same holds in this case. Indeed, it is never necessary to check these last two identities since there cannot be any cohomology when there are fewer than two $s$s. In fact, it is easy to see that there cannot be any {co}cycles at all at this level since no non-trivial combination $z_{s\psi\dots \psi}$ is annihilated by $\delta$ (provided there is at least one ${}_\psi$). We conclude that when $s<2$ (and $p>0$), $Z=\{0\}$ and, therefore, there are no new components to define nor constraints to impose. Let us refer to this as the ``2$s$s argument'' to distinguish it from the ``4$s$s'' argument given at the end of section \ref{S:1cocycle} in relation to equation (\ref{E:CohomologyRelation}).\footnote{This argument generalizes effortlessly to other superspaces. For a slightly more in-depth treatment of principal superspaces and a proof of the statement that the top two Bianchi identities impose no new constraints, see reference \cite{Randall:2014gza}.}

\subsection{Super-$3$-cocycle}
\label{S:3cocycle}
The Bianchi identities for a super-$3$-{co}cycle (\ref{E:BianchiForm}) are
\begin{align}
B_{ssss} &= 4 D_s H_{sss} + 12i H_{ss\Gamma(s,s)} \cr
B_{sss\psi} &= 3 D_s H_{ss\psi} - \partial_\psi H_{sss} - 6i H_{s\Gamma(s,s)\psi} \cr
B_{ss\psi\psi} &=  2 D_s H_{s\psi\psi} +2 \partial_\psi H_{ss\psi} +2 i H_{\Gamma(s,s)\psi\psi} \cr
B_{s\psi\psi\psi} &=  D_s H_{\psi\psi\psi} -3 \partial_\psi H_{s\psi\psi}  \cr
B_{\psi\psi\psi\psi} &= 4\partial_\psi H_{\psi\psi} .
\end{align}
We proceed with the de Rham sequence by interpreting the components of the closed $3$-{co}cycle $H$ as the obstruction to the closure of the $2$-from of the previous section. 
As the unique constraint on the superfield $W$ is the condition $D^2_{\hat a ij}W=0$ (\ref{E:Wconstraint}), we take a field $H_{\hat a ij}$ of the same form. That is, we set
\begin{align}
H_{ss\psi} = - (s^i \Sigma_{\psi}{}^{\hat a} s^j) H_{\hat a ij} .
\end{align}
Note that this component is a cocycle so that the level-0 Bianchi identity is solved with $H_{sss}=0$ (and, therefore, implies it).

With this ``initial condition'' in place, the rest of the argument is the same as in the previous two cases. The {co}boundaries at levels $\ell =1$ and $\ell=2$ give the components 
\begin{align}
H_{s\psi\psi} = \tfrac{i}{12} \epsilon_{\psi \psi}{}^{\ha \hb \hc} s^i \Sigma_{\ha \hb} D^j H_{\hc ij}
~~~\mathrm{and}~~~
H_{\psi\psi\psi} = \tfrac{1}{48} \epsilon_{\psi \psi \psi}{}^{\ha \hb} D^2_{\ha ij} H_\hb^{ij}.
\end{align}
The cohomology at these levels is given by the 4$s$s argument. For level 1, the identity (\ref{E:CohomologyRelation}) implies that (after stripping off the $s$s and $\psi$s)
\begin{align}
(\Sigma_{\hat a\hat b})_{(\hat \alpha \hat \beta} D_{\hat \gamma)(k} H^{\hat b}_{ij)} = 0 .
\end{align}
This identity is equivalent to 
\begin{align}
\left[ \delta_{\hat a}^{\hat b} \delta_{\hat \alpha}^{\hat \beta} +\tfrac15 (\Gamma_{\hat a}\Gamma^{\hat b})_{\hat \alpha}^{\hat \beta}\right] D_{\hat \gamma(k}H_{\hat b ij)} = 0,
\end{align}
where the operator in square brackets projects onto the $\Gamma$-traceless subspace. Therefore 
\begin{align}
\label{E:3formConstraint1}
C_{\hat a \hat \gamma ijk}:= D_{\hat \gamma(k}H_{\hat a ij)} - \Gamma\textrm{-trace} = 0 .
\end{align}
At level 2, the same equation is used (as always) but this time there are two $\psi$s so that the constraint is a Lorentz scalar and iso-spin triplet
\hypertarget{Cij}{}
\begin{align}
\label{E:3formConstraint2}
C_{ij} := D^2_{\hat a k(i} H_{j)}^{\hat a\, k} +6i \partial_{\hat a}H^{\hat a}_{ij} = 0.
\end{align}

\subsubsection{Independence of Constraints}
An important question in the analysis of constraints is that of their independence. In the case currently under consideration, for example, one would like to know 
whether (\ref{E:3formConstraint1}) and (\ref{E:3formConstraint2}) are compatible (in the sense that they admit non-constant solutions) and what part of the former (if any) is already implied by the latter.
For example, it is {\it a priori} possible that $C_{ij} \propto D^{k}\Gamma^{\hat a} C_{\hat a ijk}$, in which case, $C_{ij}= 0$ does not imply any new conditions, or the opposite extreme in which the two constraints together have no non-constant solutions. 

In fact, the cohomology of the algebraic complex ensures that there is no overlap at all since the image of $C_{\hat a \hat \gamma ijk}$ is in an entirely different linear subspace than that of $C_{ij}$.
In this particular example, this is expressed by the fact that $D^{k}\Gamma^{\hat a} C_{\hat a ijk}\equiv 0$ by (\ref{E:3formConstraint1}). 
Equivalently, although $D_{\hat \delta l} C_{\hat a \hat \gamma ijk}$ has a part of the form $(DC)_{\hat a ,\hat b\, ijk,l}$ and this has both a symmetric and anti-symmetric part, the $\Gamma$-tracelessness of $C_{\hat a \hat \gamma ijk}$ ensures that the symmetric part is traceless. Therefore $(DC)_{\hat a ,\hat b\, ijk,l}$ represents all the Lorentz-irreducible parts {\em except} the trace so that $D_{\hat \delta l} C_{\hat a \hat \gamma ijk}$ and $C_{ij}$ are unrelated. 

For higher cocycles and higher-dimensional constraints, the line of argument constructed directly from the superspace $D$-algebra becomes increasingly more complicated. By contrast, the homological argument is universal: Constraints arising from different levels of the Bianchi identities sit in different linear subspaces of the total cohomology $H$ and are, therefore, linearly independent. 

\subsubsection{Departure from de Rham $p$-forms}
\label{S:Orogeny}
We now come to our second surprise: The superfield $H$, derived from a $2$-form by obstructing its defining constraint, is {\em not} the super-symmetric generalization of a closed bosonic $3$-form. By the latter, we mean an irreducible superfield that contains a closed $3$-form, its superpartners, and perhaps other fields needed to complete the representation. For example, a supersymmetric version of $H_{\hat a \hat b\hat c}$ would contain the on-shell component fields $(\Phi, \psi_{\hat \alpha i} , H_{\hat a \hat b\hat c})$, perhaps together with some auxiliary fields that allow an off-shell description. There are many ways to show that the field $H_{\hat a ij}$ derived here cannot describe such an irreducible representation.\footnote{For example, because of the Lorentz index, the lowest components cannot be propagating fields and would have to be auxiliary fields. However, such superfields cannot give rise to a dynamical multiplet. We thank S. James Gates, Jr. for pointing this out.} 

What, then, is the super-3-cocycle $H_{\hat a ij}$? Since it was derived from supergeometry, one expects there to be some interpretation of such a superfield. There are (at least) two answers to this question. 
One is that composite cocycles are often of this form. This possibility is explored in some detail in reference \cite{Arias:2014ona} and we will not repeat that analysis here as much of it can be recovered from dimensional reduction (cf. \S \ref{S:RelativeCohomology}). 

Another interpretation is that it describes local superconformal gauge transformation parameters of the supergravity to which these forms couple. This interpretation was first recognized in an unrelated work \cite{Linch:2015lwa}. In the context currently under consideration, this can be seen most clearly by comparing the ``solution'' $H_{\hat a ij} = D^2_{\hat a ij} \sigma$ for an unconstrained superfield $\sigma$ to the local superconformal transformation for this superspace derived in reference \cite{Kuzenko:2007hu}. 
This interpretation ``lifts'' to six dimensions where it applies to the 4-cocycle (cf. \S \ref{S:RelativeCohomology}) (although this interpretation was not given in reference \cite{Arias:2014ona}) and seems to be the case ``generically'' in $D>4$, as we now argue.

Consider a collection of five-dimensional cocycles, the lowest non-vanishing components of which have two spinor indices: 
\begin{align}
\label{E:MissingTable}
\begin{array}{rcccccc}
\omega_{ss} 
	&=& s^2\alpha  &+& \beta_{\Gamma(s,s)} &+& \Sigma^{ab}(s^i, s^j) \gamma_{ab\, ij} \cr
\omega_{ss\psi}
	&=& (s^2 \alpha_\psi + \Gamma_\psi(s,s) \alpha^\prime) &+& \beta_{\Gamma(s,s)\psi} &+& \Sigma_\psi{}^a(s^i, s^j) \gamma_{a\, ij} \cr
\omega_{ss\psi\psi} 
	&=& (s^2 \alpha_{\psi\psi}+ \Gamma_\psi(s,s) \alpha^\prime_\psi) &+& \beta_{\Gamma(s,s)\psi \psi} &+& \Sigma_{\psi\psi} (s^i, s^j)\gamma_{ij} \cr
\omega_{ss\psi\psi\psi} 
	&=& (s^2 \alpha_{\psi\psi\psi}+ \Gamma_\psi(s,s) \alpha^\prime_{\psi\psi}) &+& \beta_{\Gamma(s,s)\psi \psi \psi} &+& \Sigma_{\psi\psi} (s^i, s^j)\gamma_{\psi\, ij} \cr
	&&&\vdots& 
\end{array}
\end{align}
Of these columns, the $\beta$s are pure gauge in the sense that there is a form $\omega^\prime$ in the same cohomology class as $\omega$ that does not have this term. (In the superspace literature, choosing $\beta = 0$ is an example of a ``conventional constraint''.)
None of the $\alpha$ terms after the first one are $\delta$-cocycles while all of the $\gamma$ terms represent cocycles except for the last one. We recognize in this table that the closed $p$-forms for low $p$ (only $p=2$ in this example) come from the $\alpha$ series whereas the closed $p$-forms for high $p$, that is, low codimension, come from the $\gamma$ series. In particular, the $4$-form is implied by the results of section \ref{S:3cocycle} to be $\gamma_{ij}$ since it is sourced by the constraint (\ref{E:3formConstraint2}). This is the linear superfield which describes the irreducible supermultiplet containing the Hodge dual of a closed codimension-1-form (see, for example, \cite{Gates:2014cqa} and references therein).

Although we are presenting this in the context of $D=5$, it is not difficult to see that this structure generically gives rise to two series of cocycles (here called $\alpha$ and $\gamma$) that have form interpretations for low and high values of $p$, respectively. When $D=4$, the end of the $\alpha$ series abuts the $\gamma$ series precisely at the crossover point, but when $D>4$ a ``gap'' opens up in which we find cocycles that do not necessarily have an interpretation as irreducible supermultiplets containing a closed $p$-form.
In sum, we have found that, in general,
\begin{align}
\begin{array}{lcr}
\boxed{\textrm{~super-de Rham $p$-{co}chain}  
	~~ \xLeftrightarrow{\hspace*{8mm}} \hspace{-8mm} \bigtimes\hspace{4mm} ~~
\textrm{supersymmetrization of de Rham $p$-form~} }
\end{array}
\nonumber
\end{align}

From the point of view of four-dimensional, $N=1$ superspace \cite{Gates:1980ay,Gates:1983nr}, this conclusion may be somewhat surprising since there is no ``gap'' in this superspace.
Nevertheless, we can recover the analogous cocycle in this complex by not assuming the vector multiplet field strength $W^\alpha$ to be chiral. That is, when we obstruct the vector multiplet Bianchi identity $\bar D_{\dot \alpha} W_\alpha - D_\alpha \bar W_{\dot \alpha} = 0$ with a superfield, we get precisely the form of the gauge transformation of the conformal graviton $\delta H_{\alpha \dot \alpha} =\bar D_{\dot \alpha} L_\alpha - D_\alpha \bar L_{\dot \alpha}$ \cite{Gates:1983nr,Buchbinder:1998qv}.\footnote{In this interpretation, the Bianchi for Bianchi identity suggests that there is a superfield $G_a$ built out of $H$ such that $G_a(\delta H)=0$. The four-dimensional, $N=1$ super-Einstein tensor is such a superfield and precisely this interpretation emerges from a reduction of a certain five-dimensional superspace \cite{Linch:2015lwa}.} 

Finally, we comment briefly on the branching and fusion of the super-de Rham complex. The results of this section can be summarized by the the diagram in figure \ref{F:Branching} representing the structure of the complex of super-$p$-cocycles.
\begin{figure}[h]
\begin{displaymath}
\xymatrix{
	&&2^\prime \ar@{->}[dr]^{\hyperlink{Caij}{C_{\hat aij}}}  & &  &   \\
0\ar@{->}[r] & 1 \ar@{->}[ur]^{\hyperlink{C}{C_{\hat a\hat bij}}} \ar@{->}[dr]_{\hyperlink{C}{C}} &&3\ar@{->}[r]^{\hyperlink{Cij}{C_{ij}}}  & 4\ar@{->}[r] &5\\
	&&2 \ar@{->}[ur]_{\hyperlink{Caij}{C_{\hat aij}}}  & & & 
 }
\end{displaymath}
\begin{caption}{\small Loops in the super-de Rham complex}
\label{F:Branching}
\small
When the reduced cohomology is reducible as a representation of the Lorentz group, branching happens in the super-de Rham complex due to the ability to source more than one constraint to generate a cocycle of in the next degree. These must eventually re-collapse by irreducibility of the cohomology in higher degree.
\end{caption}
\end{figure}

There is only a single branching and subsequent fusion in the $1\to 2\to 3$ transition. The branching is due to the fact that there were two constraints on the closed $1$-form field strength (\ref{E:1formConstraints}). The fusion is a consequence of the fact that the constraints on the field strengths of the closed $2$-form and $2^\prime$-cocycle are isomorphic as representations of the structure group: Repeating the homological analysis in the latter case implies that the $2^\prime$-cocycle is defined by a superfield of the form $W_{\hat a\hat b ij}$ subject to the dimension-2 constraint $D^{2\hat b}_{k(i} W_{\hat a \hat b j)}{}^{k} +\dots =0$ (see ref. \cite{Gates:2014cqa} for details). 
This outcome was guaranteed by the uniqueness of the cohomology found in section \ref{S:3cocycle} for the 3-cocycle.

One may be tempted to speculate on the possibility of further branching, but this is ruled out by (\ref{E:MissingTable}), interpreted now as describing the components of {\em Bianchi forms}. 
That is, branching occurs when one of the components has non-vanishing entries in both the $\alpha$ and the $\gamma$ series. In this case, it happens only in the first line, corresponding to $B_{ss} \sim s^2 C + \Sigma^{\hat a\hat b}(s^i, s^j) C_{\hat a\hat b\, ij}$. After this, only the $\gamma$ series can contribute so the rest of the complex is linear. 
Note, however, that this does not preclude the possibility of additional fusion. In order to have fusion without branching, one would need new forms that do not come from the super-de Rham complex but map into it under the action of the de Rham operator. 
We will see examples of such forms in the next section.

After this excursion into the non-$p$-form nature of certain {co}cycles of the super-de Rham complex 
(corresponding to the Lorentz non-singlets of the $\gamma$ series of (\ref{E:MissingTable})),
we now return to the question of the missing closed $3$-form.\footnote{The remaining forms in the super-de Rham complex can be found by the homological algebra argument from section \ref{S:AlgebraicComplex} with no new surprises. As our focus here is on the methodology, we present the complete results of this analysis in an accompanying work \cite{Gates:2014cqa}.}

\section{Relative Cohomology}
\label{S:RelativeCohomology}
In this section, we consider the relation of the five-dimensional, $N=1$ super-de Rham complex to that in six dimensions with $N=(1,0)$ supersymmetry \cite{Arias:2014ona}. Since our formul\ae{} of section \ref{S:General} were written without committing 
to any particular superspace, the results of that section apply equally to the six-dimensional setting. 

Again for simplicity, we specialize to flat superspace. For the purposes of exposition, let $S$ and $M$ denote the five- and six-dimensional superspaces respectively, and let $f: S\hookrightarrow M$ denote the inclusion.
We dimensionally reduce the components (\ref{E:BianchiForm}) of the Bianchi form, thought of as a six-dimensional formula on $M$, to the five-dimensional $S$. This {co}dimension-1 reduction is particularly simple because, as reviewed in appendix \ref{S:Superspace}, the $N=1$ spinor representations in five and six dimensions are isomorphic.

Let $\tilde \theta_p$ denote a closed super-$p$-form on $M$.
We pick a direction $\partial_6$ in $M$ orthogonal to $S$ and let $\theta_p:= f^\ast \tilde \theta_p$, $\tilde{\beta}_{p-1}:= \iota_{\partial_6} \tilde \theta_p$, and $\beta_{p-1}:= f^\ast \tilde \beta_{p-1}$ denote the restrictions of the components of the form to $S$. 
The reduction of the Bianchi identities is achieved by either 
({i}) wedging with $dx^6$ and then truncating to the subspace defined by $x^6=0$, or
({ii}) acting by contraction $\iota_{\partial_6}$ and then restricting. 
These cases give, respectively,
\begin{align} 
(s+1)D_s \theta_{s \dots s \psi \dots \psi} 
	-(-1)^s (p-s) \partial_\psi \theta_{s \dots s \psi \dots \psi} 
	&- i (-1)^s s(s+1) \theta_{s \dots s \Gamma(s,s)\psi \dots \psi} \cr
	&+ i (-1)^s s(s+1) c_{ss}\beta_{s \dots s\psi \dots \psi} = 0 \cr
(s+1)D_s \beta_{s \dots s \psi \dots \psi} 
	-(-1)^s (p-s) \partial_\psi \beta_{s \dots s \psi \dots \psi} 
	&- i (-1)^s s(s+1) \beta_{s \dots s \Gamma(s,s)\psi \dots \psi} =0 ,
\end{align}
where $c_{ss} = T_{ss}^6 = s^2$ denotes the contribution coming from the 6-component of the six-dimensional torsion. In de Rham notation, these read
\begin{align}
d \theta_p - c_2 \wedge \beta_{p-1} = 0
~~~\mathrm{and}~~~
d \beta_{p-1} = 0.
\end{align}

Defining the six-dimensional $(p+1)$-form $\omega_{p+1} := c_2 \wedge \tilde \beta_{p-1}$, the first equation can be rewritten as $f^\ast \omega = d \theta$. Then the pair $(\omega, \theta) \in \Omega^{p+1}(M)\times \Omega^{p}(S)$ define a $(p+1)$-cocycle in the relative de Rham complex of $S$ in $M$ with differential defined by $d(\omega, \theta):=(d\omega, f^\ast \omega - d\theta)$ \cite{bott1995differential}. Precisely this cohomology theory was used in reference \cite{Howe:2011tm} to define integration on a $D$-dimensional superspace with a $(D-1)$-dimensional boundary for $D=4$ and $D=5$. 

For our purposes (and those of ref. \cite{Howe:2011tm}), the salient feature of the relative cohomology complex is that it allows the construction of closed $p$-forms in five dimensions. 
In our case, they come from a single $p$-cocycle $\tilde \theta_p$ in six dimensions that reduces to a $p$-cochain $\theta_p$ and a $(p-1)$-cocycle $\beta_{p-1}$. Solving the condition $d\beta = 0$ as $\beta = d\alpha$ for a $(p-2)$-cochain $\alpha_{p-2}$, we obtain a five-dimensional $p$-cocycle $\theta^\prime_p$ by setting 
\begin{align}
\label{E:WeylForm}
\theta_p^\prime = \theta_p - c_2 \wedge \alpha_{p-2}
~~~\Rightarrow~~~
d\theta^\prime_p = 0.
\end{align}
The ability to construct a cocycle from the difference of two cochains in a closely related superspace was called ``Weyl triviality'' in reference \cite{Bonora:1986xd}. Here we are finding that the two required cochains exist, and have the correct property, because they descend from a single cocycle in one higher dimension. 

In the explicit $s/\psi$-component version of 
the formula (\ref{E:WeylForm}), the components of $\alpha$ generally start at a lower level ({\it i.e.} with more spinor indices) than those of $\theta$, thereby avoiding the inconsistencies in the Bianchi identities at the lowest levels for $\theta$ alone and rendering the Bianchi identities for $\theta^\prime$ consistent. Conversely, at the higher levels, this $\alpha$ correction goes to 0, not contributing to the final two components of $\theta^\prime$ (as is easily seen since these components have $s<2$). We represent this structure of $\theta^\prime$ in figure \ref{F:Ladder}. 
We now illustrate this construction explicitly in the case of the relative $3$-cocycle. 

\begin{figure}[h]
\begin{center}
\begin{tikzpicture}[rotate = 90, scale = 2]
	\draw [color = black!50!white] (-0.65, 2) -- (0.65, 2);
	\draw [color = black!50!white] (-0.65, 1.2) -- (0.65, 1.2);

	\draw (-0.5, 3) -- (-0.5, 1.85);
	\draw (-0.505, 1.785) node[right] {$\cdots$};
	\draw (-0.5, 1.35) -- (-0.5, 0.55);
	\draw (-0.505, 0.485) node[right] {$\cdots$};
	
	\draw (0.5, 3) -- (0.5, 1.85);
	\draw (0.495, 1.785) node[right] {$\cdots$};
	\draw (0.5, 1.35) -- (0.5, 0.55);
	\draw (0.495, 0.485) node[right] {$\cdots$};
	
	\draw [fill = black] (-0.5, 3) circle (0.05);
	\draw [fill = black] (-0.5, 2.5) circle (0.05);
	\draw [fill = white] (-0.5, 0.7) circle (0.05);
	\draw (-0.52, -0.125) node[right] {$\theta_p$};
	\draw (0.52, -0.125) node[right] {$c_2 \wedge \alpha_{p - 2}$};
	
	\draw [fill = white] (0.5, 3) circle (0.05);
	\draw [fill = white] (0.5, 2.5) circle (0.05);
	\draw [fill = black] (0.5, 0.7) circle (0.05);
	
	\draw [fill = black] (-0.5, 2) circle (0.05);
	\draw [fill = black] (-0.5, 1.2) circle (0.05);
	\draw [fill = black] (0.5, 2) circle (0.05);
	\draw [fill = black] (0.5, 1.2) circle (0.05);
\end{tikzpicture}
\end{center}
\begin{caption}{\small The structure of super-cocycles in relative cohomology}
\label{F:Ladder}
\small
The black nodes represent contributions to $\theta^\prime$ with the level increasing from right to left. The rungs of the ladder represent the components of $\theta^\prime$ that require terms from both $\theta$ and $c\wedge \alpha$ (represented by nodes on the respective stiles).
\end{caption}
\end{figure}

\subsection{Example: The Missing $3$-form}
In section \ref{S:3cocycle} we saw that the de Rham 3-cocycle is not the supersymmetrization of the bosonic de Rham 3-form. Here, we will construct the missing $3$-form from the relative cohomology of a super-de Rham 3-cocycle on a hypersuface in six-dimensional superspace. According to the discussion above, we reduce the six-dimensional 3-cocycle $\tilde H_3 \to H_3, F_2$ and solve $dF=0$ as $F=dA$. Then, by the usual homological argument, the $\delta$-coboundary terms give the following components for the closed 3-form $H^\prime$:
\begin{align}
\begin{array}{lcl}
H^\prime_{sss} =  - s^2 A_s  
& ~& 	
	H^\prime_{s\psi\psi} = \tfrac{i}{4} s^i \Sigma_{\psi\psi} D_i \Phi  \cr
H^\prime_{ss\psi} = \Gamma_\psi (s,s) \Phi  - s^2 A_\psi 
& ~& 
H^\prime_{\psi\psi\psi} = \tfrac{3}{16} D^i\Sigma_{\psi\psi\psi} D_i \Phi
\end{array}
\end{align}
where
$\Phi = \tfrac{i}{24} D^{\hat \alpha i}A_{\hat \alpha i}$
and 
$A_\psi = - \tfrac{i}{24} D^i \Gamma_\psi A_i$.
Similarly, the level-$\ell = 1$, $\tfrac32$, and $2$ $\delta$-cohomology imply, respectively, the constraints 
\begin{align}
\label{E:3formConstraints}
D_{(\hat \alpha (i} A_{\hat \beta) j)} = 0 
,~~~
6 (\Gamma_{\hat a} D_i)_{\hat \alpha} \Phi 
	+ 3 (\Sigma_{\hat a \hat b} D_i)_{\hat \alpha} A^{\hat b} 
	- \partial^{\hat b} (\Sigma_{\hat a \hat b} A_i)_{\hat \alpha}  = 0
,~~\mathrm{and}~~
D^2_{ij} \Phi = 0.
\end{align}
Again, we will not solve these equations here,\footnote{The first constraint was solved by Koller in six dimensions \cite{Koller:1982cs} in terms of Mezin\c{c}escu's prepotential \cite{Mezincescu:1979af}. Alternatives to this are known in harmonic \cite{Kuzenko:2005sz} and projective superspace \cite{Linch:2012zh}.} 
but we can use them to show that they give a superfield representation of the closed $3$-form.
Acting on the second constraint in (\ref{E:3formConstraints}) with a $D^{\hat \alpha}_{(j}$ and using the first condition, it follows that $D^2_{\hat a ij} \Phi = 0$, which we recognize as the defining condition (\ref{E:Wconstraint}) on the closed 2-form to which a closed 3-form is Hodge dual.

Together with the third constraint, this implies that this superfield representation of the closed $3$-form, while irreducible, is on-shell. 
One way to see this is that $D^2_{ij} W =0$ is the equation of motion of the five-dimensional vector multiplet \cite{Kuzenko:2005sz} to which the tensor is dual. Thus, a superfield satisfying both of these equations is equivalent to an on-shell vector multiplet. 
This is unsurprising considering that this representation descends from the chiral 3-form \cite{Bergshoeff:1996qm} in the six-dimensional complex, which is on-shell. (Note, however, that this is not a rule.)
An explicit component analysis confirms this interpretation \cite{Gates:2014cqa}.

We note the following features of this relative $3$-cocycle:
\begin{enumerate}
\item The natural {\it ansatz} $H_{ss\psi} = \Gamma_\psi (s,s) \Phi$ (bottom stile of figure \ref{F:Ladder}) fails to close the lowest Bianchi identity without the help of $A$ because $\Gamma_\psi (s,s)$ is not a $\delta$-cocycle in five dimensions. 
\item This {\it ansatz}, on the other hand, {does} give the correct definition of the top two components of the $3$-form. This is consistent with the fact that $A$ cannot contribute to these components since $c \wedge A$ has at least 2$s$s. 
This corresponds to the two empty nodes at the top left of figure \ref{F:Ladder}.
\item In figure \ref{F:Ladder}, there is only one rung for $p=3$ corresponding to both $\Phi$ and $A$ contributing to $H^\prime_{ss\psi}$. 
(For $p>3$ there could be more depending on the number of non-zero components in the higher-dimensional form.)
\item The lowest two components are not gauge invariant under $A_s \mapsto A_s + D_s\lambda$.
\end{enumerate}
This concludes our demonstration of the homological method for the relative de Rham cohomology. The remaining forms discovered in this way are illustrated in figure \ref{F:SubwayDiagram} where their relationship to the de Rham forms is displayed.

\begin{figure}[htb]
\begin{center}
\begin{tikzpicture}[scale = 1.8, > = {stealth}, decoration = {
    markings,
    mark = at position 0.5 with {\arrow{>}}}]
    
	\draw [postaction = {decorate}, color = Plum, thick] (-3, 0) -- (-2, 0);
	\draw [postaction = {decorate}, color = Plum, thick] (-2, 0) -- (-1, 0);
	\draw [postaction = {decorate}, color = cyan, thick, dashed] (-1, 0) -- (0, 0);
	\draw [postaction = {decorate}, color = cyan, thick] (0, 0) -- (1, 0);
	\draw [postaction = {decorate}, color = Plum, thick] (1, 0) -- (2, 0);
	\draw [postaction = {decorate}, color = Plum, thick] (2, 0) -- (3, 0);
	\draw [postaction = {decorate}, color = red, thick] (-2, 0) -- (-1, -1);
	\draw [postaction = {decorate}, color = red, thick] (-1, 0) -- (0, -1);
	\draw [postaction = {decorate}, color = red, thick] (-1, -1) -- (0, -1);
	\draw [postaction = {decorate}, color = red, thick] (0, -1) -- (1, 0);
	\draw [postaction = {decorate}, thick] (0, 1) -- (1, 0);
	\draw [postaction = {decorate}, thick] (1, 1) -- (2, 0);
	
	\draw [thick, fill = white] (-2, 0) circle (0.05);
	\draw [thick, fill = white] (-1, 0) circle (0.05);
	\draw [thick, fill = white] (0, 0) circle (0.05);
	\draw [thick, fill = white] (1, 0) circle (0.05);
	\draw [thick, fill = white] (2, 0) circle (0.05);
	\draw [thick, fill = white] (-1, -1) circle (0.05);
	\draw [thick, fill = white] (0, -1) circle (0.05);
	\draw [thick, fill = white] (0, 1) circle (0.05);
	\draw [thick, fill = white] (1, 1) circle (0.05);
	
	\draw (-3, 0) node[left] {$0$};
	\draw (3, 0) node[right] {$0$};
	\draw (-2, 0.1) node[above] {$A$};
	\draw (-1, 0.1) node[above] {$W$};
	\draw (0, 0.1) node[above] {$\Phi$};
	\draw (1, 0.1) node[above] {$G$};
	\draw (2, 0.1) node[above] {$K$};
	\draw (-1, -0.9) node[above] {$X$};
	\draw (0, -0.9) node[above] {$H$};
	\draw (0, 1.1) node[above] {$Y$};
	\draw (1, 1.1) node[above] {$Z$};
\end{tikzpicture}
\end{center}
\begin{caption}{\small Topology of the five-dimensional superform ``complex''}
\label{F:SubwayDiagram}
\small
The top and bottom rows consist of the (non-matter) relative cohomology forms and the (non-matter) super-de Rham forms, respectively, whereas the matter multiplets have been arranged to lie on the middle row.
The solid lines denote the action of the super-de Rham differential $d$. (The dashed line represents an unknown map; the other unknown maps have been omitted.)
The blue lines indicate that these forms result from supersymmetrizing the bosonic de Rham complex, while the red lines trace the super-de Rham complex. (Purple lines are both.)
We have also included the additional forms not otherwise mentioned: the alternative 2-form $X_{\hat{a} \hat{b} ij}$, the alternative-and-relative 3-form $Y_{\hat{a} \hat{b} ij}$, and the relative 4-form $Z_{\hat{a} ij}$.
(It turns out that the would-be relative 2-form is equivalent (up to zero mode) to the super-de Rham 2-form $W$ so it has been dropped from the top row.)
\end{caption}
\end{figure}

\section{Outlook}
\label{S:Outlook}

In an effort to understand the structure of differential forms in superspace, we have investigated the super-de Rham complex of cocycles in five dimensions and its relation to the analogous complex in six dimensions. Among the things we have learned is that the cocycles we need are to be found in the Chevalley-Eilenberg complex of the supersymmetry algebra with values in superfields and that their components can be computed with minimal effort from that same complex with constant coefficients. The cohomology of the latter generally reduces to a few (one in 5D and two in 6D) non-trivial terms that determine the structure of the entire complex. In the cases we considered, this structure branched and fused creating loops in what is usually a linear chain complex. 

We have also learned that such cocycles generally fail to be supersymmetric $p$-forms in the sense that they do not describe irreducible supermultiplets containing a closed form of degree $p$. This knowledge is prerequisite to the construction of dynamical theories in superspace. For example, it is clear now that, were one to attempt to describe the gauge theory of a dynamical 2-form, one should take as a starting point the closed 3-form of the relative super-de Rham complex rather than the 3-cocycle of the de Rham complex. This insight also leads to the reinterpretation of the 3-cocycle as conformal supergravity gauge transformation parameters or, possibly, composite forms needed to preserve the DG-algebra structure as was found to be the case in six dimensions \cite{Arias:2014ona}.
Finally, we gained insight into ``ectoplasm with an edge'' constructions \cite{Howe:2011tm} and the higher-dimensional origin of Weyl triviality \cite{Bonora:1986xd}.

There are many directions in which to expand this line of investigation, of which we mention two. Firstly, there is the extension to superspaces of dimension other than $5$ and $6$. For these applications, the superspaces for which the vector $\gamma^a(s,s)$ is null have the simplest cohomological structure, as is suggested by the exposition in appendix \ref{S:Superspace}. These famously correspond to the spaces with $D=2+2^k$ for $k=0,1,2, \dots$ with the missing cases gotten by dimensional reduction on the cohomology with constant coefficients. 

Secondly, we would like to generalize the construction to curved superspace by coupling to conformal supergravity. Happily, this too requires relatively minor changes to the framework mostly having to do with the inclusion of additional torsions and the corrections to the superfield constraints they imply. (See ref. \cite{Arias:2014ona} for the six-dimensional curved space analogue of the super-de Rham complex.) In fact, the flat-space cohomology already determines part of the structure of the supergravity torsions, as suggested by our observations regarding the 3-cocycle and its relation to local superconformal gauge transformations. 
Work is currently underway to use this observation to determine the supergravity torsions (and thus the supergeometry, cf. \cite{Dragon:1978nf}) through their couplings to forms thereby circumventing the usual analysis of curved superspace Bianchi identites. 


\section*{Acknowledgements}
W{\sc dl}3 thanks the participants of the UMD RIT on Geometry and Physics for the many discussions from which this work has benefited greatly. Special thanks are due to S. James Gates Jr., Paul Green, and Richard Wentworth for their insights, encouragement, and many hours of stimulating discussion.
W{\sc dl}3 also thanks the Simons Center for Geometry and Physics for hospitality during the {\sc xii} Simons Workshop.

This work is supported in part by National Science Foundation grants
PHY-0652983 and 
PHY-0354401
and the University of Maryland Center for String \& Particle Theory. 
SR was also supported by the Maryland Summer Scholars program and the Davis Foundation.

\appendix 
\section{Five- and Six-dimensional Superspace}
\label{S:Superspace}
In this appendix, we derive the properties of five- and six-dimensional superspaces with eight supercharges needed to determine the structure of de Rham cocycles in the main part of the text. These properties can be copied directly from references \cite{Arias:2014ona} and \cite{Kuzenko:2005sz}, but we rederive them here in a form that generalizes easily to other superspaces.

Although we will mostly be focusing on five dimensions, it is easiest to derive the properties of this superspace from the six-dimensional superspace in which it embeds. 
We take the spinor representation in six dimensions to be pseudo-Majorana with index structure $s^{\alpha i}$, where $\alpha = 1, \dots, 4$ and $i=1,2$ are $SL(4;\mathbb R)$ ``spin'' and $SU(2)$ ``iso-spin'' indices, respectively. The off-diagonal blocks $(\gamma_a)_{\alpha\beta}$ and $(\tilde \gamma_a)^{\alpha \beta}:= \tfrac12 \varepsilon^{\alpha \beta\gamma \delta}(\gamma_a)_{\gamma \delta} $ of the Dirac matrices (Pauli matrices) are antisymmetric in their spinor indices. In terms of these, the Clifford algebra rules reduce to the form
\begin{align}
\gamma_a \tilde \gamma_b = - \eta_{ab} + \gamma_{ab}.
\end{align}
This defines $\gamma_{ab} := \gamma_{[a} \tilde \gamma_{b]} = - \tilde \gamma_{[a} \gamma_{b]} =: -\tilde \gamma_{ab}$.
A commuting spinor $s^{\alpha i}$, defines a vector and a triplet of self-dual 3-planes by the combinations
\begin{align}
\gamma_a(s,s) := s^{\alpha i} (\gamma_a)_{\alpha \beta} s^\beta_i
~~~\mathrm{and}~~~
\gamma_{abc} (s^i,s^j) := s^{\alpha i} (\gamma_{abc})_{\alpha \beta} s^{\beta j} 
	=\gamma_{abc} (s^j,s^i),
\end{align}
where $\gamma_{abc}:=\gamma_{[a} \tilde \gamma_b \gamma_{c]}$ denotes the anti-symmetric part of the product of three Pauli matrices. The vector defined by $s$ is null
\begin{align}
\label{E:Principal}
\gamma^a(s,s) \gamma_a(s,s)  = 0.
\end{align}
Useful Fierz identities can be derived by polarizing $s \to s+t$ and expanding in powers of $t$. For example, the first such relation implies the famous identity 
\begin{align}
\label{E:Principal_Pol}
\gamma^a(s,s) \gamma_a(s,t)  = 0
\end{align}
for all commuting spinors $s$ and $t$.\footnote{
This relation is generally valid for $D= 2^k+2$ for $k=0,1,2,3$ when the spinor structure is Majorana, Weyl, pseudo-Majorana, and Majorana-Weyl, respectively \cite{Kugo:1982bn}.
As it and its consequences are the only non-trivial relations we use regarding the (iso-)spin structure of the superspace, the cohomological analysis we perform should be extendible to these cases with minimal modifications.
} 
Setting 
$t^i= \omega^{ij}_{ab}\tilde \gamma^{ab}s_j$ for some $\omega^{ij}_{ab}$
and substituting, we obtain 
\begin{align}
\left[
	\gamma_{[a}(s,s) \gamma_{b]}(s^i,s^j) 
	+ \gamma^{c}(s,s) (s^i\gamma_{abc}s^j) 
\right] \omega_{ij}^{ab}	
= 0 .
\end{align}
Since $\omega$ is arbitrary, the expression in brackets must vanish. The first term vanishes irrespective of the symmetry properties of the iso-spin indices on $\omega$. The second term is non-trivial only if $\omega$ has a symmetric part. Therefore, we find that the vector defined by $s$ is ``orthogonal'' to the triplet of self-dual 3-planes it defines:
\begin{align}
\label{E:Orthogonality}
\gamma^c(s,s) \gamma_{abc} (s^i,s^j) = 0.
\end{align}

Finally, we will need the completeness relation. 
Such a relation is equivalent to the statement that the Dirac matrices (or the Pauli matrices $\gamma$ and $\tilde \gamma$) generate the Clifford algebra and, as such, does not correspond to additional information that is put in ``by hand''. However, as we have chosen to follow the conventions of reference \cite{Linch:2012zh}, we are no longer free to normalize the independent terms which come out to be
\begin{align}
\label{E:Completeness}
X_s Y_s = 
	\tfrac{1}{8} \gamma^a(s,s)\tilde \gamma_a(X,Y) 
	+ \tfrac{1}{8\cdot 3!} \gamma^{abc}(s^i,s^j) \tilde \gamma_{abc} (X_i ,Y_j),
\end{align}
for any two co-spinors $X_{\alpha i}$ and $Y_{\alpha i}$. 
Here $8$ is the number of real supercharges in this superspace.

With the relations (\ref{E:Principal_Pol}) and (\ref{E:Orthogonality}) and the normalizations (\ref{E:Completeness}), we are ready to reduce to five dimensions. 
The spinor representation stays pseudo-Majorana with the difference that the spinor indices can be raised and lowered with the $Sp(4;\mathbb R)$ invariant proportional to $(\gamma_6)_{\alpha \beta}$. 
Distinguishing five-dimensional indices with a caret where necessary, we define 
$(\Gamma_{\hat a})_{\hat \alpha}{}^{\hat \beta} = (\gamma_{\hat a 6})_{\hat \alpha}{}^{\hat \beta}$ and 
$(\Sigma_{\hat a \hat b})_{\hat \alpha}{}^{\hat \beta} = \tfrac12(\gamma_{\hat a \hat b 6})_{\hat \alpha}{}^{\hat \beta}$ (so chosen to agree with the conventions of \cite{Kuzenko:2005sz,Linch:2012zh}). With this understood, the equations (\ref{E:Principal_Pol}) and (\ref{E:Orthogonality}) reduce to
\begin{align}
\label{E:5DCohomologyRelations}
\Gamma^a(s,s)\Gamma_a(s,t) + s^2 t_s = 0 
~~~\mathrm{and}~~~
\Gamma^\ha(s, s) \Sigma_{\ha \hb}(s^i, s^j) = 0 .
\end{align}
It will be important to our analysis in section \ref{S:5DN1deRham} that the five-dimensional vector defined by the commuting spinor $s$ is no longer null but that it is ``orthogonal'' to the triplet of 2-planes defined by it.
The completeness relation (\ref{E:Completeness}) becomes
\begin{align}
\label{E:5DCompleteness}
X_s Y_s = 
	\tfrac{1}{8} s^2 X^iY_i
	- \tfrac{1}{8} (X^i \Gamma_{\Gamma(s,s)}Y_i) 
	+ \tfrac{1}{2} \Sigma^{\hat a \hat b}(s^i,s^j) (X_i \Sigma_{\hat a\hat b} Y_j).
\end{align}
All of the relative coefficients in the formul\ae{} we derive in the main text for the components of the cocycles and the constraints on their defining superfields are simple combinations of only 
the universal coefficients in the Bianchi forms
and the relative coefficients in these five-dimensional Fierz identities.


{\small
\bibliography{/Users/wdlinch3/Dropbox/Rashoumon/LaTeX/BibTex/BibTex}
\bibliographystyle{unsrt}
}

\end{document}